\documentstyle[12pt,epsf,worldsci]{article}

\begin{document}
\begin{flushright}
JHU--TIPAC--940016
\end{flushright}

\title{Possible Puzzles in Nonleptonic $B$ Decays}

\author{Adam F.~Falk
\thanks{Invited talk presented at the Eighteenth Johns Hopkins Workshop on
Current Problems in Particle Theory, Florence, Italy, August 31--September 2,
1994.}
\\
{\it Department of Physics and Astronomy}\\
{\it The Johns Hopkins University}\\
{\it 3400 North Charles Street}\\
{\it Baltimore, Maryland 21218 U.S.A.}}

\maketitle

\begin{abstract}
I discuss the recent controversy over the semileptonic branching ratio of the
$B$ meson, pointing out that it is only the combination of this quantity with
the reported charm multiplicity in $B$ decays which is in serious conflict with
theoretical calculations.  I consider possible solutions to the problem.
\end{abstract}

\section{Introduction}

There has been considerable recent interest in the question of whether there is
a serious conflict between the measured semileptonic branching ratio of the $B$
meson and its theoretical calculation~\cite{BBSV,FWD}.  While a straightforward
analysis of this quantity, based on the operator product expansion, certainly
yields a result which is in conflict with experiment, is this discrepancy worth
worrying about?  In particular, is it pointing us toward new physics beyond the
Standard Model, or telling us something interesting about QCD, or neither?
What is the most conservative, and hence most plausible, explanation of the
puzzle?  Here I will argue that its resolution lies most likely within QCD
itself and is probably quite mundane, namely that the calculation of inclusive
nonleptonic decay rates is considerably less trustworthy than the calculation
of inclusive semileptonic decays and fails even for total energy releases as
large as $m_b$.

I will concentrate on the theoretical foundation of the calculation, explaining
the roles which are played by the Operator Product Expansion (OPE) and the two
versions of the duality hypothesis, ``local'' and ``global''.  We will see that
if we combine the OPE and the assumption of local duality with the reported
semileptonic branching ratio of the $B$ meson, then we are led to the
expectation of excess charm production in inclusive $B$ decays, as much as 1.3
charm quarks per decay.  The charm data, as reported, do not support such an
excess; if anything, they indicate a small {\em deficit\/} of charm.  What is
more, the most plausible and least ``damaging'' emendations to the OPE do not
solve this problem, as they also would imply excess charm.  The real puzzle is
not the semileptonic branching ratio alone, but rather the combination of it
with the reported charm multiplicity.  We are led to the conclusion that, if
the data are confirmed, the key assumption of local duality may not be
justified in nonleptonic $B$ decays.

\section{Theoretical Analysis of Inclusive $B$ Decays}

The inclusive decay rate of the $B$ meson may be organized by the flavor
quantum numbers of the final state,
\begin{equation}\label{totalrate}
   \Gamma_{{\rm TOT}} = \Gamma(b\to c\,\ell\bar\nu) + \Gamma(b\to
   c\bar ud') + \Gamma(b\to c\bar cs')\,.
\end{equation}
We neglect rare processes, such as those mediated by an underlying
$b\to u$ transition or penguin-induced decays.  We denote by $d'$ and $s'$ the
approximate flavor eigenstates $(d'=d\cos\theta_1-s\sin\theta_1$,
$s'=d\sin\theta_1+s\cos\theta_1)$ which couple to $u$ and $c$,
respectively, and we ignore the effect of the strange quark mass.  It is
convenient to normalize the inclusive partial rates to the
semielectronic rate, defining
\begin{equation}
   R_{ud} = {\Gamma(b\to c\bar ud')\over
             3\Gamma(b\to c\,{\rm e}\bar\nu)}\,,\qquad\qquad
   R_{cs} = {\Gamma(b\to c\bar cs')\over
             3\Gamma(b\to c\,{\rm e}\bar\nu)}\,.
\end{equation}
The full semileptonic width may be written in terms of the
semielectronic width as
\begin{equation}
   \Gamma(b\to c\,\ell\bar\nu) = 3f(m_\tau)
   \Gamma(b\to c\,{\rm e}\bar\nu)\,,
\end{equation}
where the factor $3f(m_\tau)$ accounts for the three flavors of
lepton, with a phase space suppression which takes into account the
$\tau$ mass.  Then, since the semileptonic branching ratio is given by
$Br(b\to c\,\ell\bar\nu) = \Gamma(b\to c\,\ell\bar\nu)/
\Gamma_{{\rm TOT}}$, we may rewrite Eq.~(\ref{totalrate}) in the form
\begin{equation}\label{rsum}
   R_{ud} + R_{cs} = f(m_\tau)\,{1-Br(b\to c\,\ell\bar\nu)
   \over Br(b\to c\,\ell\bar\nu)}\,.
\end{equation}
The measured partial semileptonic branching fractions are~\cite{pdb,aleph}
\begin{eqnarray}
   Br(\overline B\to X{\rm e}\bar\nu) &=& 10.7\pm0.5\%\,,\cr
   Br(\overline B\to X\mu\bar\nu) &=& 10.3\pm0.5\%\,,\cr
   Br(\overline B\to X\tau\bar\nu) &=& 2.8\pm0.6\%\,,
\end{eqnarray}
leading to a total semileptonic branching fraction $Br(b \rightarrow
c\,\ell\bar\nu)$  of $23.8\pm0.9\%$, with the experimental errors added
in quadrature.  Of the semileptonic rate, $11\%$ comes from decays to
$\tau$, corresponding to a phase space suppression factor $f(m_\tau)=0.74$,
consistent with what one would expect in free quark
decay~\cite{tau,CPT}. If we substitute the measured branching fractions
into the right-hand side of Eq.~(\ref{rsum}), we find
\begin{equation}\label{rsumdata}
   R_{ud} + R_{cs} = 2.37\pm0.12\,.
\end{equation}
This is the result which we must try to reproduce theoretically.

To compute $R_{ud}$ and $R_{cs}$, it is necessary to compute individually the
partial widths $\Gamma(b\to c\,e\bar\nu)$, $\Gamma(b\to c\bar ud')$ and
$\Gamma(b\to c\bar cs')$.  Because the computation is conceptually more
straightforward for this case, we begin with the nonleptonic widths.

\subsection{Nonleptonic decays}

The nonleptonic quark decay $b\to c\,\bar q_1q_2$ is mediated by four-quark
operators of the form
\begin{equation}\label{operator}
   {\cal O} = \bar c \Gamma_1 b\,\bar q_2\Gamma_2 q_1\,,
\end{equation}
where the $\Gamma_i$ denote generic Dirac matrices.  Such operators are induced
by the exchange of $W$ bosons which have been integrated out the theory at the
scale $\mu=M_W$, where $\Gamma_1=\Gamma_2=\gamma_\mu(1-\gamma_5)$ and the
coefficient of $\cal O$ in the effective Hamiltonian is $G_F/\sqrt2$.  As the
theory is evolved down to the scale $\mu=m_b$, this coefficient changes, and
additional operators of the form (\ref{operator}) are induced with coefficients
which depend on $\log(\mu/m_b)$.  In this schematic discussion of the inclusive
decay rate, we will ignore the interference between these operators and treat
only the generic operator $\cal O$ above.

The inclusive decay rate of the $\overline B$ meson mediated by the operator
$\cal O$ is given by
\begin{equation}
   \Gamma_{\cal O} \sim \sum_X\langle\overline B|\,{\cal O}^\dagger\,|X\rangle
   \langle X|\,{\cal O}\,|\overline B\rangle\
   = \sum_X\big|\langle X|\,{\cal O}\,|\overline B\rangle\big|^2\,,
\end{equation}
where the sum runs over all states $X$ with the appropriate quantum numbers.
This expression may be rewritten {\it via\/} the optical theorem in terms of
the imaginary part of a forward scattering amplitude,
\begin{equation}
   \Gamma_{\cal O} \sim {\rm Im}\,\langle \overline B|\,T\{
   {\cal O}^\dagger,{\cal O}\}\,
   |\overline B\rangle\,.
\end{equation}
The next step is to compute the time-ordered product $T\{{\cal O}^\dagger,{\cal
O}\}$ in perturbative QCD.

\begin{figure}
  \epsfxsize=12cm
  \centerline{\epsfbox{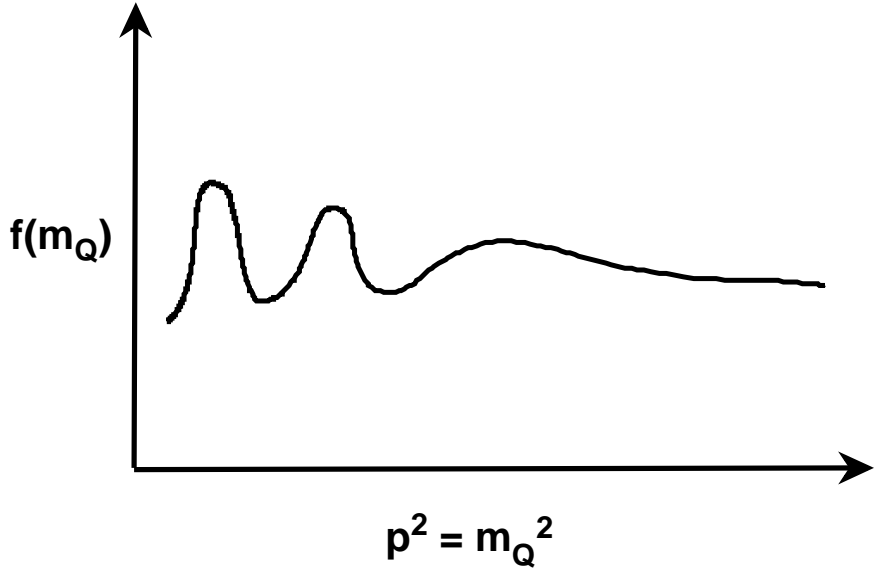}}
  \caption{Cartoon of $f(m_Q)\equiv{\rm Im}\,\langle M_Q|\,T\{{\cal
            O}^\dagger,{\cal O}\}\,|M_Q\rangle$.}
  \label{cartoon}
\end{figure}

Before we proceed with this step, it is useful to examine the conditions under
which it is sensible to apply a perturbative analysis.  In Fig.~\ref{cartoon},
I present a cartoon of the quantity
\begin{equation}
   f(m_Q)\equiv{\rm Im}\,\langle M_Q|\,T\{{\cal
   O}^\dagger,{\cal O}\}\,|M_Q\rangle\,,
\end{equation}
where the $b$ quark has been replaced with a quark $Q$ of mass $m_Q$ and the
$\overline B$ meson by the ground state meson $M_Q$.
The expectation value of the correlator is given as a function of $m_Q$.  We
see in Fig.~\ref{cartoon} that the behavior of $f(m_Q)$ is qualitatively
different for $m_Q$ large and for $m_Q$ small compared to the scale
$\Lambda_{\rm QCD}$ which characterizes the nonperturbative regime of QCD.  For
small $m_Q$, the correlator is dominated by a few resonances whose position and
strength depends on nonperturbative strong dynamics.  In this ``resonance
regime'', no perturbative calculation can reproduce the rapid variations in the
shape of $f(m_Q)$.  For large $m_Q$, however, these rapid variations damp out,
as in this region the correlator is dominated by multiparticle states.  While
even here there are new thresholds associated with the production of additional
pions, their effect is small compared to the smooth background of states to
which they are being added.  In this regime, one might expect the smooth output
of a perturbative calculation of the correlator to approach the physical
$f(m_Q)$.  This property, that at large enough energies one may compute
accurately within perturbative QCD, is known as ``local duality''.

Where is the transition between the resonance and perturbative regimes, or in
other words, above what value of $m_Q$ does one expect local duality to hold?
In particular, is $m_b$ sufficiently large for perturbative QCD to be
applicable?  Since we cannot vary $m_b$ experimentally, we cannot answer this
question by tracing out $f(m_Q)$ and looking for where it becomes smooth.  We
can only argue on dimensional grounds.  Na\"\i vely, we might expect to be well
in the perturbative regime,  since $m_b\approx5\,{\rm GeV}$ is large compared
to typical QCD scales of hundreds of MeV.  However, the situation is not
necessarily so clear, since the nonleptonic decay is into three colored
particles which must divide the available energy between them, and charm
quark(s) in the final state carry away substantial rest energy as well.  We
also note that the transition point between the resonance and perturbative
regimes undoubtedly depends on the operator $\cal O$, as well as on the
external states $M_Q$.

While we are free to have our prejudices, it is probably best to leave this
question open for the time being.  We may plunge ahead with the perturbative
calculation of the total width, while bearing in mind that the assumption of
local duality is made by necessity but cannot rigorously be justified.  Later,
it will be necessary to evaluate any discrepancies between experiment and the
theoretical results in light of their sensitivity to this crucial assumption.

It is also true that local duality may well fail at different $p^2$ in the
channels $b\to c\bar ud'$ and $b\to c\bar cs'$.  In fact, we might expect it to
fail first (at higher energy) in the channel with two charm quarks in the final
state, since, due to the charm quark rest energy, the kinetic energy released
per particle is lower in this channel.  For example, in a simple free quark
decay model the $s$ quark has an average energy of only 1 GeV.  This intuitive
expectation is our first hint, with more to come, that the quantity $R_{ud}$ is
more reliably calculable than $R_{cs}$.

\subsection{Semileptonic decays}

The situation is considerably better for semileptonic decays.  Here the decay
is mediated by a product of currents,
\begin{equation}
   {\cal O} = \bar c\gamma^\mu(1-\gamma^5)b\,\bar\ell\gamma_\mu(1-\gamma^5)
   \nu\,,
\end{equation}
and the matrix element factorizes into hadronic and leptonic pieces,
\begin{equation}
   \langle X\,\ell(p_\ell)\bar\nu(p_{\bar\nu})|\,
   J^\mu_h J_{\ell\mu}\,|\overline B\rangle = \langle X|\,J^\mu_h\,|
   \overline B\rangle\cdot
   \langle \ell(p_\ell)\bar\nu(p_{\bar\nu})|\,J_{\ell\mu}
   |0\rangle\,\,.
\end{equation}
We then find an expression in which the integral over the momenta of the
leptons is explicit,
\begin{equation}\label{lepint}
   \Gamma\sim\int{\rm d}y\,{\rm d}v\cdot\hat q\,{\rm d}\hat q^2
   \,L_{\mu\nu}(v\cdot\hat q,\hat q^2,y)\, W^{\mu\nu}(v\cdot\hat q,\hat
   q^2)\, ,
\end{equation}
where $L_{\mu\nu}$ is the lepton tensor and $W^{\mu\nu}$ the hadron
tensor. Here the momentum of the external $b$ quark is written as
$p_b^\mu=m_b v^\mu$. The other independent kinematic variables are
$q^\mu=p^\mu_\ell+p^\mu_{\bar\nu}$ and $y=2v\cdot p_\ell/m_b$.  It is
convenient
to scale all momenta by $m_b$, so $\hat q=q/m_b$.  The hadronic tensor
is given by
\begin{eqnarray}
   W^{\mu\nu} &=& \sum_X \langle \overline B|\,J_h^{\mu\dagger}\,|X\rangle
                  \langle X|\,J_h^\nu\,|\overline B\rangle \cr
              &=& -2\,{\rm Im}\,\langle \overline B|\,i\int{\rm d}x\,
                  e^{iq\cdot x}
                  \,T\left\{J_h^{\mu\dagger}(x),J_h^\nu(0)\right\}
                  \,|\overline B\rangle\cr
              &\equiv& -2\,{\rm Im}\,T^{\mu\nu}\,.
\end{eqnarray}
We may perform the integrals in $y$, $v\cdot\hat q$ and $\hat q^2$ in
Eq.~(\ref{lepint}) to compute the total semileptonic decay rate, or
leave some of them unintegrated to obtain various differential
distributions.

\begin{figure}
  \epsfxsize=12cm
  \centerline{\epsfbox{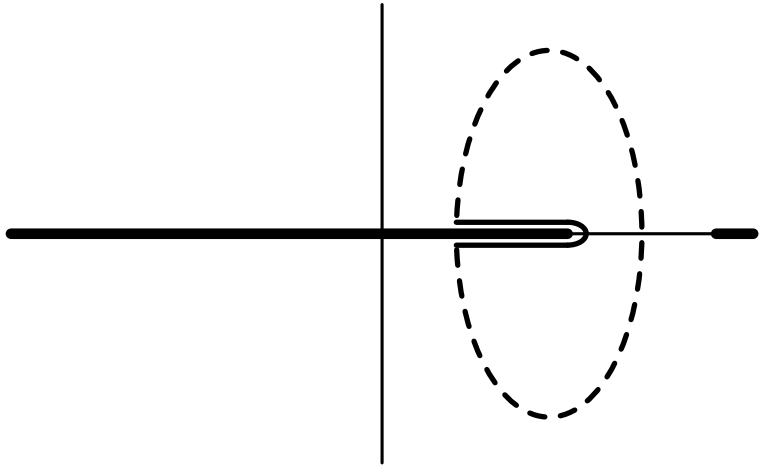}}
  \caption{Cuts and contours in the complex $v\cdot\hat q$ plane, for fixed
  $\hat q^2$ and $y$.  The cut extends along the real axis except in the region
  $(1+\hat q^2-\hat m_q^2)/2< v\cdot\hat q<((2+\hat m_q)^2-\hat q^2-1)/2$.  The
  endpoints of the contour integral are at
  $v\cdot\hat q=(y+\hat q^2/y)/2\pm i\epsilon$.}
  \label{contours}
\end{figure}

Let us consider as an example the doubly differential distribution ${\rm
d}\Gamma/{\rm d}y\,{\rm d}\hat q^2$, for which we must perform the integral
over $v\cdot\hat q$, for $y$ and $\hat q^2$ fixed.  The range of integration is
given by $(y+\hat q^2/y)/2\le v\cdot\hat q\le (1+\hat q^2-\hat m_q^2)/2$, where
$m_q$ is the mass of the quark to which the $b$ decays, and $\hat m_q=m_q/m_b$.
 The integration is pictured as the solid contour in Fig.~\ref{contours}, along
with the analytic structure of $T^{\mu\nu}$; the cut extends along the real
axis except in the region $(1+\hat q^2-\hat m_q^2)/2< v\cdot\hat q<((2+\hat
m_q)^2-\hat q^2-1)/2$.
Note that we have already included only the imaginary part of $T^{\mu\nu}$ by
integrating over the top of the cut and then back underneath.  In general,
along the cut $T^{\mu\nu}$ will look qualitatively like the cartoon shown in
Fig.~\ref{cartoon}, depending on $v\cdot\hat q$ in a complicated
nonperturbative way.  However, we may now use Cauchy's Theorem to deform the
contour of integration to the dashed contour in Fig.~\ref{contours}, which lies
far away from the cut everywhere but at its endpoints.  Along this new contour,
we are far from any physical intermediate states, and it is safe to perform the
operator product expansion in perturbative QCD.

The additional integrals over the lepton phase space allow us to determine the
physical rate in terms of certain smooth integrals of $T^{\mu\nu}$ rather than
in terms of the time-ordered product evaluated at a single point.  These smooth
integrals may be computed reliably by deforming the contour into the unphysical
region.  That we we can compute integrals of $T^{\mu\nu}$ in this way is the
property of ``global duality''.  In principle, it rests on a much firmer
theoretical foundation than does local duality, which asserts the computability
of QCD correlators directly on the physical cut rather than off in some
unphysical region of the complex plane.

Unfortunately, the dashed contour must still approach the physical cut near the
endpoints of the integration, introducing an unavoidable uncertainty into the
calculation.  There is a two-pronged argument that this uncertainty is small.
First, for large $m_b$, the proportion of the contour which is within
$\Lambda_{\rm QCD}$ of the physical cut scales as $\Lambda_{\rm QCD}/m_b$ and
thus makes a small contribution to the total integral.  Second, in this small
region one may invoke local duality, as in the case of nonleptonic decays.  The
endpoints of the integration lie conveniently at the point of maximum recoil of
the final state quark, where this property is most likely to hold. In addition,
since there are fewer colored particles in the final state than in nonleptonic
decays, one would expect this assumption to be more reasonable in the
semileptonic than in the nonleptonic case.

The point is that semileptonic decays, because they rely on global rather than
local duality, may be calculated much more reliably in perturbative QCD than
may nonleptonic decays.  This is why we have chosen, in defining $R_{ud}$ and
$R_{cs}$, to normalize the nonleptonic widths to the semielectronic width.  (Of
course, this also eliminates the dependence on the CKM parameter $V_{cb}$ and
some of the dependence on $m_b$.)  The theoretical uncertainties in $R_{ud}$
and $R_{cs}$ will be dominated entirely by the uncertainties in the numerators.

\subsection{The perturbative calculation}

\begin{figure}
  \epsfxsize=12cm
  \centerline{\epsfbox{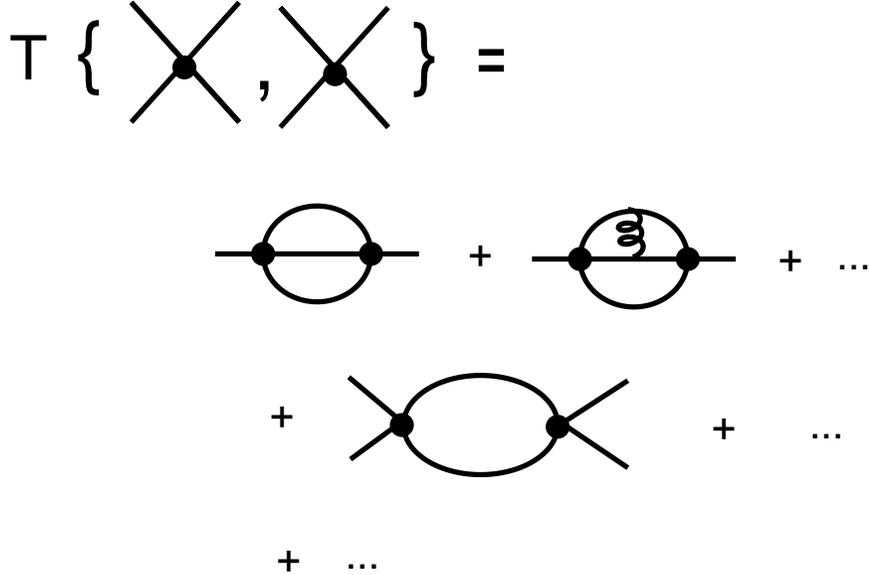}}
  \caption{The operator product expansion of $T\{{\cal O}^\dagger,
  {\cal O}\}$.}
  \label{OPE}
\end{figure}

The computation of $T\{{\cal O}^\dagger,{\cal O}\}$ in perturbation theory
follows well-known techniques for both semileptonic~\cite{CGG,BSV,MW,TM,FLS}
and  nonleptonic~\cite{BS} decays.  Since the quarks in the intermediate state
are assumed to be far from the mass shell by an amount of order $m_b$, we may
develop the operator product in a simultaneous expansion in $1/m_b$ and
$\alpha_s$.  Graphically, the expansion proceeds as in Fig.~\ref{OPE} in the
nonleptonic case, to whatever accuracy is desired.
The Operator Product Expansion takes the form of an infinite series of local
operators,
\begin{equation}
   T\{{\cal O}^\dagger,{\cal O}\} = \sum_{n=0} {c_n(\alpha_s)\over m_b^{n+1}}
   {\cal O}_n\,,
\end{equation}
where the operator ${\cal O}_n$ has dimension $n+3$ and the $c_n$ are
calculable functions of $\alpha_s$.  The notation is somewhat schematic, in
that more than one operator may appear for a given $n$.  For example, the
operators of lowest dimension which appear in the expansion are of the form
\begin{eqnarray}
   {\cal O}_0 &=& \bar b\Gamma b\,,\nonumber\cr
   {\cal O}_1 &=& \bar b\Gamma^\mu D_\mu b\,,\cr
   {\cal O}_2 &=& \bar b\Gamma^{\mu\nu} D_\mu D_\nu b\,,\nonumber
\end{eqnarray}
where $D_\mu$ is the covariant derivative, and $\Gamma^{\mu\dots}$ are
combinations of Dirac matrices.

Once the OPE has been performed, we must compute the expectation values of the
local operators ${\cal O}_n$ in the $\overline B$ state.  Such expectation
values involve the structure of the meson as determined by nonperturbative QCD,
and typically are incalculable.  However, here we may use the additional
symmetries of the heavy quark limit~\cite{HQET} to simplify the description of
these matrix elements.  In this limit, we construct an effective theory in
which the production of heavy quark-antiquark pairs is eliminated, at the price
of introducing into the Lagrangian an infinite tower of nonrenormalizable
operators suppressed by powers of $1/m_b$.  In the effective theory, we may
compute the matrix elements which arise at a given order in $n$ in terms of a
few nonperturbative parameters.  In fact, the matrix element of ${\cal O}_0$ is
determined exactly by the conservation of heavy quark number; if the expansion
is truncated at this point, then we reproduce precisely the result of the free
quark decay model.  Since all matrix elements of ${\cal O}_1$
vanish~\cite{CGG}, the leading corrections to free quark decay come only at
relative order $1/m_b^2$, not at order $1/m_b$.  These corrections come from
initial state binding effects of the $b$ quark in the $\overline B$ meson, the
energy of the heavy quark being altered from its ``free'' value by its
interactions with the cloud of light quarks and gluons which surrounds it.
They are governed by only two parameters, defined by the expectation values
\begin{eqnarray}
   &&\langle\overline B|\,\bar b(iD)^2b\,|\overline B\rangle
   \equiv 2M_B\lambda_1\,,\cr
   &&\langle\overline B|\,\bar b(-\textstyle{i\over2}\sigma^{\mu\nu})G_{\mu\nu}
   \,|\overline B\rangle\equiv 6M_B\lambda_2\,.
\end{eqnarray}
The parameter $\lambda_2$, associated with an operator which violates the heavy
quark spin symmetry~\cite{HQET}, is determined by the mass difference between
the $B$ and $B^*$ mesons,
\begin{equation}
   \lambda_2={1\over2}(M_{B^*}^2-M_B^2)\approx0.12\,{\rm GeV}^2\,.
\end{equation}
Unfortunately, $\lambda_1$ cannot at this time be measured experimentally, and
for its estimation we must rely on models of QCD.  The same is true of the
matrix elements of operators of dimension six and higher.

The ratios $R_{ud}$ and $R_{cs}$ receive contributions, then, both from
perturbative ($\alpha_s$) and nonperturbative ($\lambda_1$, $\lambda_2,\ldots$)
sources.  I will summarize below the analysis of the semileptonic~\cite{JK,YN}
and nonleptonic~\cite{GLAM,AP,HP,BBBG} radiative corrections which exists in
the literature.  The nonperturbative corrections~\cite{BSV,MW,TM,BS} to
$R_{ud}$ and $R_{cs}$ have recently been studied up to operators of dimension
six and shown to be quite small~\cite{BBSV}; because they turn out to be
unimportant compared to the uncertainties in the radiative corrections, I will
not include them from here on.

\begin{figure}
  \epsfxsize=12cm
  \centerline{\epsfbox{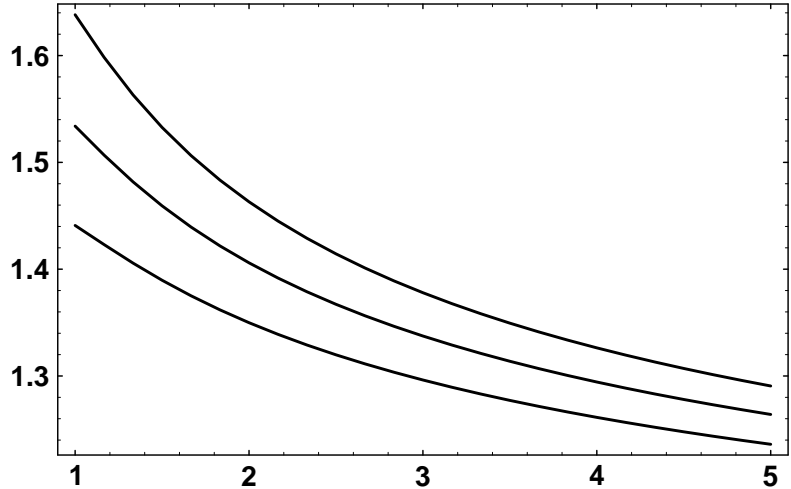}}
  \caption{The radiative correction $P(\mu)$.  The upper curve
  corresponds to $\Lambda^{(5)}_{\overline{{\rm MS}}}=220\,{\rm MeV}$, the
  middle curve to $\Lambda^{(5)}_{\overline{{\rm MS}}}=180\,{\rm MeV}$,
  and the lower curve to
  $\Lambda^{(5)}_{\overline{{\rm MS}}}=140\,{\rm MeV}$.  We take
  $m_b=4.8\,{\rm GeV}$.}
  \label{Pofmu}
\end{figure}

After some manipulation~\cite{FWD}, the ratios $R_{ud}$ and $R_{cs}$ may be
written in the form
\begin{eqnarray}\label{rbars}
   R_{ud} &=& P(\mu)+\delta P_{ud}(\mu,m_c)\,,\cr
   R_{cs} &=& G(m_c/m_b)\bigg[P(\mu)+\delta P_{cs}(\mu,m_c)\bigg]\,.
\end{eqnarray}
Here
\begin{equation}
   P(\mu)=\eta(\mu)\left[1+{\alpha_s(\mu)\over\pi}+J_2(\mu)\right]
\end{equation}
is the radiative correction in the massless ($m_c=0$) limit, with leading
logarithmic~\cite{GLAM} ($\eta(\mu)$), one loop~\cite{JK} ($\alpha_s(\mu)/\pi$)
and subleading logarithmic~\cite{AP} ($J_2(\mu)$) contributions.  The inclusion
of the subleading logarithms softens the dependence of $P(\mu)$ on the
renormalization scale, which is shown in Fig.~\ref{Pofmu} for a range of values
of $\Lambda^{(5)}_{\overline{{\rm MS}}}$.
The quantities $\delta P_{ud}$ and $\delta P_{cs}$ parameterize the effect on
the radiative corrections of the charm quark mass.  Recent
calculations~\cite{FWD,JK,YN,BBBG} show $\delta P_{ud}$ to be no larger than
$\sim0.02$.  There are indications~\cite{HP,BBBG,MV} that $\delta P_{cs}$ may
be substantially larger than $\delta P_{ud}$, but a full calculation is not yet
available; we consider the uncertainty due to $\delta P_{cs}$ below.

\begin{figure}
  \epsfxsize=12cm
  \centerline{\epsfbox{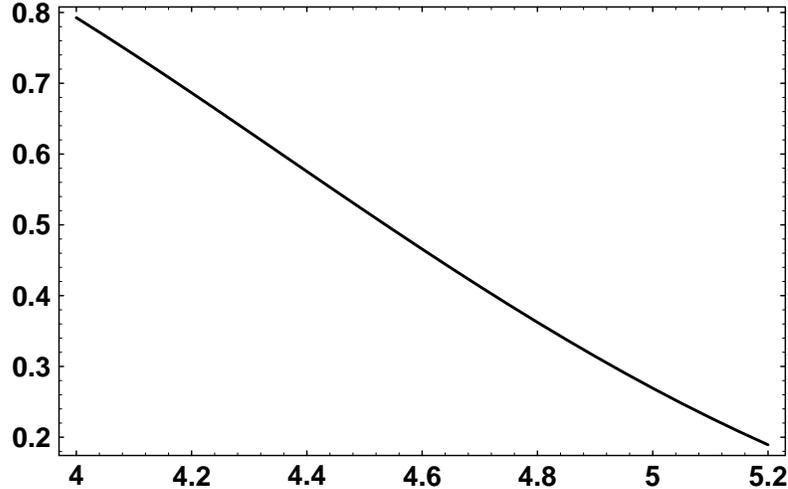}}
  \caption{The phase space suppression factor $G(m_c/m_b)$, as a
  function of $m_b$ with $m_b-m_c=3.34\,{\rm GeV}$ held fixed.}
  \label{Gofmc}
\end{figure}

While there is certainly some variation in $R_{ud}$ and $R_{cs}$ due to the
dependence of the radiative corrections on the renormalization scale, the
largest uncertainty in $R_{cs}$ comes from the phase space factor $G(m_c/m_b)$.
There is an additional dependence on the charm mass in this channel because of
the second charm quark in the final state.  The analytic expression for
$G(m_c/m_b)$ has been computed~\cite{CPT}; we present a plot in
Fig.~\ref{Gofmc}.
Here we have used the fact that within the heavy quark expansion, the
difference between $m_c$ and $m_b$ is much more precisely known than either of
the masses individually, and is given in terms of the spin-averaged $D$ meson
and $B$ meson masses:
\begin{equation}\label{massdiff}
   m_b-m_c=\langle M_B\rangle_{{\rm ave.}} -
           \langle M_D\rangle_{{\rm ave.}} = 3.34\,{\rm GeV}\,.
\end{equation}
In Fig.~\ref{Gofmc} we hold $m_b-m_c$ fixed and consider variations of $m_b$
only.

The important inputs into the theoretical calculation, then, are $\mu$, $m_b$
and, to a lesser extent, $\Lambda^{(5)}_{\overline{{\rm MS}}}$.  As first
emphasized by Bigi et al.~\cite{BBSV}, it is not a trivial task to fix these
parameters to satisfy the experimental constraint $R_{ud} + R_{cs} =
2.37\pm0.12$.  For example, if one takes the reasonable values
$\mu=m_b=4.8\,{\rm GeV}$ and $\Lambda^{(5)}_{\overline{{\rm MS}}}=180\,{\rm
MeV}$ and neglects $\delta P_{ud}$ and $\delta P_{cs}$, then $P(\mu)=1.27$,
$G(m_c/m_b)=0.36$ and $R_{ud}+R_{cs}$ is only 1.73, far short of the
experimental number.  The question, then, is whether there exist values of the
inputs for which we can manage to reproduce the data.

It is useful to start by considering $R_{ud}$, for which the calculation is
likely to be more reliable.  To enhance this quantity we may vary the
renormalization scale $\mu$.  The choice $\mu=m_b$ is motivated by the fact
that the total energy released in the decay $b\to c\bar ud'$ is $m_b$; however,
since this energy has to be divided among three colored particles, perhaps the
appropriate scale is lower.\footnote{This scenario is supported by a recent
estimate~\cite{LSW} of $\mu$ for semileptonic decays using the BLM
scale-setting criterion~\cite{BLM}, in which a scale $\mu$ as much as an order
of magnitude below $m_b$ is indicated.}  For $\mu=1.6\,{\rm GeV}\approx m_b/3$,
for example, and $\Lambda^{(5)}_{\overline{{\rm MS}}}=220\,{\rm MeV}$, we have
$R_{ud}=P(\mu)=1.52$, a modest enhancement.  If we take this value of $\mu$ as
a reasonable lower limit, then the data on the semileptonic branching ratio
implies $R_{cs}\ge0.85$.  This restriction on $R_{cs}$ follows simply from the
assumption that the perturbative calculation of $R_{ud}$ is reliable.

There is more than one way to achieve $R_{cs}\ge0.85$, both within the present
calculation and by going outside it.  The easiest, perhaps, is to lower $m_b$
to $4.4\,$GeV, which corresponds to $m_c=1.1\,$GeV.  These masses for the
bottom and charm quarks are somewhat low, but probably not unacceptably so.
Taking $\mu=1.6\,$GeV, then, we would find $G(m_c/m_b)=0.58$, and
$R_{cs}=0.89$.  Alternatively, we note that recent
calculations~\cite{HP,BBBG,MV} indicate that $\delta P_{cs}$ may not be
negligible at all, and that the inclusion of charm mass effects in the
radiative correction in this channel may lead to an enhancement of as much as
thirty percent~\cite{MV}.  Finally, we could postulate the failure of local
duality in the $b\to c\bar cs'$ channel, since this is the channel in which we
expect this assumption to be the least reliable.  From any of these points of
view, or from a combination of them, the $B$ semileptonic branching ratio is
surprising, perhaps, but certainly explicable.  We simply have to allow for an
enhancement of the final state with two charmed quarks relative to the final
state with one, as compared to our na\"\i ve expectation.  At the same time,
such an enhancement is {\it absolutely necessary\/} if the data are to be
reconciled with the theoretical prediction for $R_{ud}$.

\section{Experimental consequences}

This proposal that an enhancement of $R_{cs}$ explains the observed
semileptonic branching ratio is not without experimental implications of its
own.  The most striking of these is a large number of charmed quarks per $B$
decay.  The charm multiplicity $n_c$ may be written
\begin{equation}\label{nc}
   n_c = 1 + R_{cs}\, {Br (\overline B \to X_c \ell\bar \nu)\over
   f(m_\tau)}\,.
\end{equation}
Using $Br
(\overline B\to X_c\,\ell\bar\nu) = 23.8\%$ and $f(m_\tau) = 0.74$ in
Eq.~(\ref{nc}) yields
\begin{equation}
   n_c = 1.00 + 0.32\,R_{cs}\,,
\end{equation}
which for the values of $R_{cs}$ necessary to explain the semileptonic
branching ratio would indicate $n_c\sim 1.3$.

While there has been no direct inclusive measurement of $n_c$, it may be
estimated by summing individual partial widths.  Among the $B$ decay products,
there are contributions to $n_c$ from charmed mesons, charmed baryons, and
$c\bar c$ resonances.  The number of charged and neutral $D$ mesons per decay,
summed over $B$ and $\overline B$, has been measured to be~\cite{BHP}
\begin{eqnarray}\label{dmulti}
   n_{D^\pm} &=& 0.246\pm0.031\pm0.025\,,\cr
   n_{D^0,\overline D^0} &=& 0.567\pm0.040\pm0.023\,.
\end{eqnarray}
The branching ratio to $D_s^\pm$ mesons has not yet been determined,
because no absolute $D_s$ branching ratio has been measured.
However, it is known that~\cite{CLEO1}
\begin{equation}\label{dsdata}
   n_{D^{\pm}_s} = (0.1181 \pm0.0053\pm0.0094)
   \left[{3.7\%}\over Br (D_s\to\phi \pi)\right]\,,
\end{equation}
and the branching ratio for $D_s\to\phi\pi$ is expected to be about
$3.7\%$.  Inclusive $B$ decays to $c\bar c$ resonances below $D\overline D$
threshold contribute two units to $n_c$.  The branching ratio to $\psi$ has
been measured to be $(1.11\pm 0.08)\%$, including feed-down from the excited
states $\psi'$ and $\chi_c$~\cite{BHP}.  Combined with the
measurements~\cite{pdb} $Br(B\to\psi'X)=(0.32\pm0.05)\%$, $Br(B\to\chi_{c1}
X)=(0.66\pm0.20)\%$ and $Br(B\to\eta_c X)<1\%$, we expect that the inclusive
$B$ branching ratio to charmonium states below $D\overline D$ threshold is
between two and three percent.

Finally, there is the contribution to $n_c$ of $B$ decays to charmed baryons.
While the production of charmed baryons may be estimated from the measured
inclusive production of all baryons in $B$ decays~\cite{BHP}, such an
extraction requires strong assumptions about the underlying production process
and the predominant decay chains~\cite{DCFW}.  Such assumptions introduce
considerable uncertainty.  Recently, however, the CLEO Collaboration has
reported certain exclusive branching ratios of $B$ mesons to charmed
baryons~\cite{CLEO2}:
\begin{eqnarray}\label{xifrac}
   n_{\Xi_c^+}&\sim&1.5\pm0.7\%\,,\cr
   n_{\Xi_c^0}&\sim&2.4\pm1.3\%\,,
\end{eqnarray}
with a branching ratio to $\Lambda_c$ roughly twice that to $\Xi_c$.  Only
statistical errors are included here.  (The new CLEO data do not support the
recent suggestion~\cite{DCFW} that charmed baryon production in $B$ decays is
dominated by the underlying process $b\to c\bar cs'$, which would yield final
states with two charmed baryons rather than one, somewhat alleviating the
``charm deficit''.)

Summing over all the contributions, then, the experimental charm multiplicity
is $n_c\approx1.1\pm0.1$, where the error is dominated by the uncertainty on
charmed baryon production.  Such a value does not provide evidence for the
excess charm production indicated by the data on the semileptonic branching
ratio.

\section{Theoretical implications}

What are we to make of this situation?  We have seen that given the
uncertainties in the calculation, neither the semileptonic branching ratio of
the $B$ meson nor the charm multiplicity in $B$ decays is by itself much of a
problem.  However, the combination of the two measurements is in sharp
contradiction with current theoretical calculations.  While there is
considerable room in the computation of either quantity to make it agree with
experiment, the adjustments that make one quantity better make the other worse.
 When viewed in this light, there are important implications for the theory of
inclusive $B$ decays.

Of course, it is certainly possible that in the future the data may move in the
direction of the theoretical calculations, and that the discrepancy simply will
go away.  In particular, this may happen once the contributions to $n_c$ of
charmed baryons is better understood.  If future measurements give support to
the prediction $n_c\sim1.3$, then the resolution of the semileptonic branching
ratio puzzle simply will be that $R_{cs}$ is larger than expected.  Whether
this is due to a failure of local duality, to a low bottom quark mass, or to a
renormalization scale much lower than the na\"\i ve choice $\mu=m_b$, may still
be unclear, but as we have stressed, such an enhancement is not hard to arrange
theoretically.

If, on the other hand, the present measurement of $n_c$ is confirmed, then the
consequences are more extreme.  The most natural explanation of such a
situation would be that, for some reason, the calculation of $R_{ud}$ is also
not reliable.  Again, this could be due simply to a failure of local duality:
perhaps $m_b$ is not large enough to calculate any inclusive nonleptonic decay
widths at all.  Such a conclusion would be disappointing, of course, since this
is the nonleptonic channel in which local duality was considered more likely to
hold.  However, disappointing is not the same as unnatural, and it is
worthwhile to recall that the invocation of local duality was based essentially
only on ``educated hope''.  There is nothing profound, necessarily, in learning
that this time our luck was not so good.

An alternative explanation would be that the initial
decomposition~(\ref{totalrate}) of the total width was incomplete, in that
charmless final states were neglected.  A charmless branching ratio of $20\%$
would be sufficient to solve the semileptonic branching ratio puzzle without
inducing an unacceptable charm multiplicity.  Of course, the neglect of
charmless final states is well justified within the Standard Model, in which
$b\to u$ weak transitions are suppressed relative to $b\to c$ by a factor
$|V_{ub}/V_{cb}|^2\sim0.01$, and transitions such as $b\to sg$ and $b\to
s\gamma$ are found at the $10^{-3}$ level or below.  However, in extensions of
the Standard Model this need not be the case, and it is possible to construct
scenarios~\cite{AK} in which, for example, the rate for $b\to sg$ is
substantially enhanced without inducing dangerous new contributions to $b\to
s\gamma$, for which there are already stringent limits.  Models with an
enhancement of $b\to sg$ are also constrained by the known limits on the
exclusive decay $B\to K\pi$, although unknown strong interaction effects
preclude one from translating these limits into a firm bound on the strength of
the $b\to sg$ coupling.

Although it is certainly exciting to consider such novel possibilities, I must
stress that they are not as yet well motivated by the data.  The most
conservative, natural and plausible resolution of the semileptonic branching
ratio puzzle lies within QCD itself.  In fact, the entire subject ought best to
be thought of as an interesting test of the applicability of the Operator
Product Expansion and perturbative techniques to the calculation of inclusive
nonleptonic decays in the $B$ system.  As of today, the jury remains out on
this important question.

\bigskip
\noindent{\bf Acknowledgments}
\medskip

It is a pleasure to thank the organizers of the Workshop for putting together
such an enjoyable and interesting conference.  This work was supported by the
National Science Foundation under National Young Investigator Award PHY-9457916
and Grant No.~PHY-9404057, and by the Department of Energy under Outstanding
Junior Investigator Award DE-FG02-094ER40869.

\end{document}